# Phase Structure of Resource Allocation Games

Robert Savit*, Sven A. Brueckner[+], H.Van Dyke Parunak[+] and John Sauter[+]

*Corresponding author, Department of Physics
University of Michigan, Ann Arbor, MI, 48109
[+]ALTARUM, PO Box 134001, Ann Arbor, MI 48113-4001

## Abstract

We consider a class of games that are generalizations of the minority game, in that the demand and supply of the resource are specified independently. This allows us to study systems in which agents compete for a resource under different demand loads. Among other features, we find the existence of a robust phase change with a coexistence region as the demand load is varied, separating regions with nearly balanced supply and demand from regions of scarce or abundant resources. The coexistence region exists when the amount of information used by the agents to make their choices is greater than a critical value which is related to the point at which there is a phase transition in the standard minority game.

Competition for resources is ubiquitous in social and biological systems. Animals foraging for food, companies competing for market share, clients competing for bandwidth on the Internet and politicians competing for votes are just a few examples.[1] In at least some such systems, agents making choices that differ from most of their competitors can lead to increased benefit for the agent. One important attempt to abstract and model the dynamics of being different is the minority game,[2] which has a remarkable phase structure as a function of the amount of information that agents use to make their choices.

While it clearly captures some important dynamics in competition for limited resources, the minority game, as it is usually specified, is limited to a specific ratio of supply and demand for the resource. By construction, at most (N-1)/2 of the N agents playing the minority game can be rewarded in a given time step of the game. In many real systems, the supply/demand may be quite different than in the minority game, and so it is of considerable interest to study games in which this ratio can have different values from that of the standard minority game. In this paper we present a class of such models, of which the minority game is a special case, and study the way in which system behavior differs for different loads (demand vs. supply). In particular, we show that there is a phase change as the relative demand on the resource changes, and that the phase diagram includes a coexistence region in which the collection of games with the same control parameters bifurcates into two distinct groups with very different behaviors.

Consider, a game in which N agents compete for a resource from one of two suppliers. We will consider the games with more than two suppliers elsewhere.[3] At each time step of the game, each supplier has available C/2 units of the resource, and each agent chooses one of the two suppliers as a source for one unit of the resource. The agents will make their choices of which supplier to choose at a given time step, using a mechanism similar to that used in the minority game. In particular, each agent is endowed with s (in the games considered here, s=2) strategies. Each strategy is a look up table in which data from a set of publicly available information is used as input to determine an agent's decision. The set of publicly available information is the historical time series of which of the suppliers had more requests for resource than that supply could satisfy, as a function of time. Let $n_j(t)$ be the number of agents requesting resource from supplier j at time t. Supplier j is underloaded at time t if $n_j(t) \leq C/2$, and is overloaded otherwise. We indicate underloading of a supplier by +, and overloading by -. Then, the state of the system at any given time is defined by a two-tuple (a,b), where a indicates the state (over- or under-loaded) of supplier 0 and b indicates the state of supplier 1. Although there are, in principle, four possible states, the number of accessible states depends on the relative values of N and C. If $N \leq C$, then states (+,+), (+,-) and (-,+) are possible. If N > C + 1, states (-,-), (+,-) and (-,+) are possible. For the special case N=C+1, this game reduces to the minority game and only states (+,-) and (-,+) are accessible. Thus, if the agents use information in their strategies from the last m time steps of the game, the dimension of the strategy space will be $3^m$, except if N=C+1, in which case the dimension of the strategy space will be $2^m$.

An agent must choose which of its two strategies to play at a given time. Following the scheme of the standard minority game, an agent will choose to play that strategy which would, up to that point in the game, have been responsible for the greatest gain for the agent, had that strategy been played for all past times of the game. Thus, the relative ranking of an agent's strategies will depend on the payoff to the agents. In this letter, we will consider games with two different payoff schemes. The first, called binary payoff, awards one point to each agent using an underloaded supplier, while agents using an overloaded supplier get nothing. These same awards are made to strategies to determine their relative rankings. The second payoff scheme, called partial satisfaction, awards one point to each agent using an underloaded supplier, while each agent using an overloaded supplier is awarded a fraction of a point equal to C/2n, where n (>C/2) is the total number of agents using that supplier at that time step. The same scheme is used to award agents' strategies. Specifically, the strategy of an agent that is actually played is awarded the same points as the agent. To evaluate a strategy not played, one awards one point to that strategy if it would have chosen an underloaded supplier at some time step, and, awards C/(2n) points if it would have chosen an overloaded supplier, where n is the number of agents actually using the overloaded supplier at that time step. Note that these awards are made assuming the same distribution of agents among the suppliers as actually occurred. No correction is made for the fact that had the unplayed strategy been played, the distribution of agents might have differed by one. Thus, this strategy ranking scheme is similar to that of the naïve agents used in the first studies of the minority game. Most of the results reported in this paper are for games played with binary satisfaction. However, the most important aspect of our results are robust when the payoff scheme is that of partial satisfaction, as we shall explain below.

Let $\sigma$ be the standard deviation of $n_1(t)$ averaged over time for one game. The general behavior of this set of games is illustrated in Figure 1 in which we present a plot of the average value of $\sigma^2/N$ as a function of N and C, averaged over 13 different runs for each value of N and C for binary satisfaction.[4] The same plot for games played with partial satisfaction is qualitatively similar. All games in this plot were played with agents all of whom had strategies that used information from the last 4 time steps of the game—i.e, m=4. Plots for different values of m, while differing in important ways (some of which are described here, and some described elsewhere[4]) have the same general structure. At the extremities, (large N, small C and large C small N) are two regions in which $\sigma^2/N$ is fairly smooth as a function of N and C. As we move in toward the diagonal, (N=C+1), we pass into areas in which the dependence of $\sigma^2/N$ on C and N is rougher. Moving further toward the diagonal from either direction, $\sigma^2/N$ decreases and, at least for m>2, has a smoother dependence on C and N. Finally, very close to the diagonal $\sigma^2/N$ increases, reaching its maximum value near C=N.

This figure has many very interesting features that will be elucidated in detail elsewhere.[4] In this letter we want to focus primarily on the most general overall structure, and, in particular, on the obvious difference between the region near the diagonal, in which $\sigma^2/N$ has a local maximum, and the regions further from the diagonal.

To understand qualitatively what is going on, it is useful to look at a typical sample of the time series of, say, $n_1(t)$ for a game in the region of the central peak, and for a game from a region further from the diagonal. Typical examples are shown in Fig. 2. To facilitate comparison both these games have N>C. In these figures, the dashed lines indicate the range of values inside of which the system is in the state (-,-), while outside the dashed lines the system is in the state (-,+) or (+,-). Note that in Fig.2a the system is almost always in either the state (-,+) or (+,-) (about 90% of the time), while for Fig. 2b, the system is almost always in the state (-,-). This is significant, since in both cases, all three states (-,-), (-,+) and (+,-) are in principle accessible to the system. In the special case of the minority game, with N=C+1, the system must be in either (-,+) or (+,-) at each time step, by construction. However, it is clear from Fig. 2a, that games played with other configurations of N and C not too far from the minority game configuration are dynamically driven to behavior which appears to be similar to that of the minority game. On the other hand, if N is too large for a given C, the system is in a much different phase, one dominated by (-,-) states in which agents typically are not rewarded. We call the region in which the system is dominated by (-,+) or (+,-) states, the region of limited resources, while the region in which the system is dominated by the state (-,-) is the region of scarce resources. The region away from the central peak, but with N<C, is dominated by the state (+,+), and we refer to that as the region of abundant resources.

That games with N>>C (N<<C) should be dominated by (-,-) ((+,+)) states is not surprising. It is also not unreasonable to suppose that configurations near the minority game should be dominated by (-,+) or (+,-) states (although *quantitatively* the minority game does differ dramatically from its neighbors, as we shall describe below). But what is noteworthy and surprising is the way in which system behavior changes as we move from the region of limited resources to either scarce or abundant resources.

To see this, refer to Fig. 3. In this figure, we plot $\sigma^2/N$ for each of 32 runs for a range of values of N (N>C) with C=200 and m=6. We see very clearly a region in N in which different games segregate into one of two bands. An examination of the time series of $n_1(t)$ for these runs indicates that those in the upper band are qualitatively similar to Fig. 2a, being dominated by the states (-,+) and (+,-), while those in the lower band are qualitatively similar to Fig. 2b, dominated by the state (-,-). The upper band is a smooth continuation of the central peak region seen, for example, in Fig. 1, while the lower band smoothly continues to the area of scarce resources, also seen in Fig. 1. Thus, in this example, the transition between limited and scarce resources proceeds through a coexistence region in which the collection of games, all with the same control parameters (m, N and C), but with different assignments of initial strategies dynamically bifurcates into two distinct groups. The first group exhibits dynamics similar to that seen in the standard minority game, while the second group exhibits dynamics that are quite different. The behavior of the system in the scarce (or abundant) resource phase has some interesting features and will be discussed in detail elsewhere.[4]

The bifurcated coexistence behavior as shown in Fig. 3 exists for values of m greater than a certain minimum, $m_c^*$. If m is too small, the clear bifurcated coexistence disappears and is replaced by a broad, smooth, but noisy crossover between the limited resource region (in which the system dynamics are like those of the minority game) and the scarce (or, if N<C, abundant) resource region. The value of m, $m_c^*$, below which there is no coexistence region is very significant. To understand its significance, look at Fig. 4, in which we plot $\sigma^2/N$ as a function of m for both the minority game, and for neighboring games with N=C and N=C+2. We see here similar curves, but with the phase transition offset. The simple reason is that for N≠C+1, the dimension of the strategy space for the games is $3^m$ rather than $2^m$ as it is for the minority game. It turns out that the bifurcated coexistence region exists only for $m>m_c^*$, the point of the phase transition for non-minority game configurations. To avoid the singular distinction between the case N=C+1 and N≠C+1, it is convenient to consider the behavior of games as a function of $\delta \equiv D/N$, where D is the dimension of the strategy space, i.e. $D=2^m$ for the usual minority game, and $D=3^m$ for the case N≠C+1. Plotted as a function of this variable, the dips in Fig. 4 coincide at a value $\delta_c \sim 1/3$. Thus, the bifurcated coexistence requires a per capita amount of information equal to that that occurs at the phase transition in usual minority game.

Parenthetically, and as we shall discuss in detail elsewhere,[5] we comment that if the strategy space is sampled non-uniformly, there are considerable complications that arise both in the standard minority game and in the more general resource allocation games that we discuss here. In such a case, it is tempting to consider the behavior of these games as a function of dynamically generated variables, rather than as a function of external control variables, such as N, C and m. Among the dynamically generated variables that most strongly suggest themselves is $\zeta \equiv e^S/N$, where S is the entropy associated with the string of m-tuples that constitute the publicly available information. In the limit that all allowed m-tuples appear with equal probability, this quantity reduces to $\delta$. Using $\zeta$ as a variable rather than $\delta$ is illuminating, but carries with it it's own complications, particularly vis-à-vis the problem of scaling in these games. This will be discussed in considerable detail elsewhere.[5]

Finally, we note that for large m the bifurcated coexistence region persists, even though *within* each game the agents' strategy choices are largely random. For large m in the coexistence region, $\sigma^2/N$ either has the value ¼, associated with minority game-like behavior (dominated by system states (+,-) and (-,+)), or 1/8, reflecting the dynamics typical of the scarce resource region (dominated by system states (-,-)). Intermediate values do not exist.[4]

These results are summarized in Fig. 5, in which we present a qualitative phase diagram for this system. The vertical axis represents "load" on the system. In the case of experiments with fixed C, this can be thought of as a monotonic function of N. The horizontal axis carries a measure of the normalized (per capita) information used by the agents to make their decisions. In this figure we use $\delta$. This figure should be understood to be only qualitative. Determining the precise positions of the phase boundaries goes well beyond the scope of this work. However, this figure does capture the following important features: 1. The region of limited resources, dominated by minority game-like dynamics is qualitatively distinct from the regions of scarce or abundant resources[4] and 2. For values of $\delta > \delta_c$, there is a coexistence region as we move from limited to scarce or abundant resources. In this region, the collection of games bifurcates, so that each game takes on features either of a system in the limited or in the scarce (or abundant) resource region. Games with intermediate behavior do not exist. Finally, we have observed that the position of the bifurcated coexistence region varies in an interesting way with the size of the system. In particular, for $\delta \approx \delta_c$, and for a given C, the value of N at which one observes bifurcated coexistence, $N_b$, satisfies $N_b-C \propto C^p$, where $1/2 \geq p \geq 3/4$.[4]

It is also important to point out that the bifurcated coexistence region is robust to some significant changes in the game. In particular there continues to be a bifurcated coexistence region when games are played with partial satisfaction rather than binary satisfaction. This is very important, since it suggests that, like the phase transition in the minority game, the coexistence region may be a universal feature, mediating a transition between two phases in a large class of games.

In this paper we have examined an important class of resource allocation games that are generalizations of the minority game. As the demand load on the system varies away from the minority game configuration (N=C+1) the system continues to exhibit minority-game like behavior until the demand is sufficiently high (or low). At that point the system exhibits a transition from a region of competition for limited resources (minority game-like behavior) to one of competition for scarce (or abundant) resources. If $\delta \geq \delta_c$ the transition between these qualitatively different states is mediated through a surprising bifurcated coexistence region. We have also studied resource allocation systems with more than two suppliers. The general phase structure we have found here applies in those cases also, but is somewhat more complicated.[3]

Based on our work, several important questions suggest themselves. First, it is clear that in the coexistence region the initial distribution of strategies to the agents strongly affects which branch a given game will occupy. However, the initial distribution of strategies to the agents is not always determinative of which branch a given game will occupy in the coexistence region. This will be discussed in more detail elsewhere.[4] Second, it is unclear what the fate of the bifurcation phenomenon is upon the introduction of evolution for the strategies. Third, our analysis has made no direct allusion to agent wealth. There is an interesting interpretation of agent wealth in the coexistence region, and that also will be discussed elsewhere.[4] Finally, our work illustrates the continuing, even growing importance of the application of concepts from the physical sciences to problems of collections of adaptive agents in the social and biological sciences.

This work was supported in part by the DARPA ANTS program under contract F30602-99-C-0202 to ERIM CEC, under DARPA PM's Janos Sztipanovits and Vijay Raghavan and Rome COTR Dan Daskiewich. The views and conclusions in this document are those of the authors and should not be interpreted as representing the official policies, either expressed or implied, of the Defense Advanced Research Projects Agency or the US Government.

---

[1] Throughout this paper, we use "compete" and "competition" informally. Our agents do not explicitly seek to frustrate one another's goals, but simply behave based on their own self-interest. In some contexts, this may be an important distinction. For a more detailed discussion, see H.V.D. Parunak, S. Brueckner, M. Fleischer, and J. Odell, in preparation.

[2] D. Challet and Y.-C. Zhang, *Physica A*, **246**, 407 (1997); R. Savit, R. Manuca and R. Riolo, Phys. Rev. Lett. **82**, 2203 (1999); R. Manuca, Y. Li, R. Riolo and R. Savit, *Physica A,* **282** 559-608 (2000); D. Challet, M. Marsili and R. Zecchina, Phys. Rev. Lett. 84, 1824 (2000. For further references to the extensive literature on this game, see the excellent web site http://www.unifr.ch/econophysics/minority.

[3] R. Savit, S. Brueckner, H. V. D. Parunak and J. Sauter*, Resource Allocation Games with More Than Two Suppliers*. in preparation.

[4] A larger range of C is plotted to show a symmetry about the minority game diagonal N=C+1, that exists for games with binary satisfaction. See R. Savit, S. Brueckner, H. V. D. Parunak and J. Sauter*, General Structure of Resource Allocation Games*, to appear for more details.

[5] R. Savit, Y. Li, S. Brueckner, H. V. D. Parunak and J. Sauter, in preparation. The question of nonuniform sampling of the strategy space has also been briefly touched on in D. Challet and M. Marsili, preprint, cond-mat/0004196 v1 (revised Oct. 2001).

**Figure Captions**

Fig. 1 $\sigma^2/N$ as a function of N ($3 \leq N \leq 50$) and C ($2 \leq C \leq 70$) for m=4.

Fig. 2. Segments of the time series of $n_1(t)$ for a games played in two different regions. Values of $n_1(t)$ that place the system in the states (+,-), (-,-) or (-,+) are indicated by the dashed lines. a.) N=203, C=200, m=6, the limited resource region, b.) N=209, C=200, m=6, the scarce resource region.

Fig. 3. $\sigma^2/N$ for different runs as a function of N. ($201 \leq N \leq 210$) for C=200 and m=6. 32 runs are presented for each value of N. This figure illustrates the bifurcated coexistence region.

Fig. 4. $\sigma^2/N$ as a function of m for N=C+1 (the minority game configuration) and N=C and N=C+2, neighboring the minority game configuration.

Fig. 5. A qualitative phase diagram for the class of game discussed here. The vertical axis is load on the system, and the horizontal axis is a measure of normalized information used by the agents to make their choices, specifically $\delta \equiv D/N$.

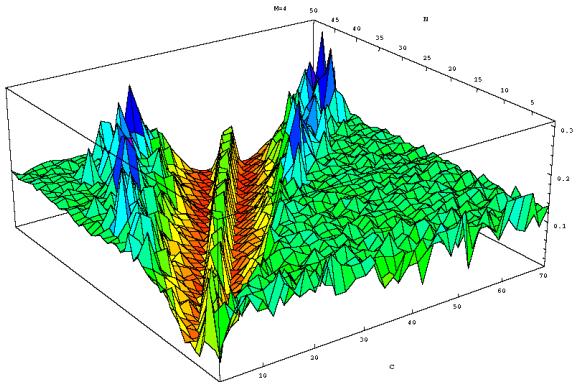

**Figure 1**

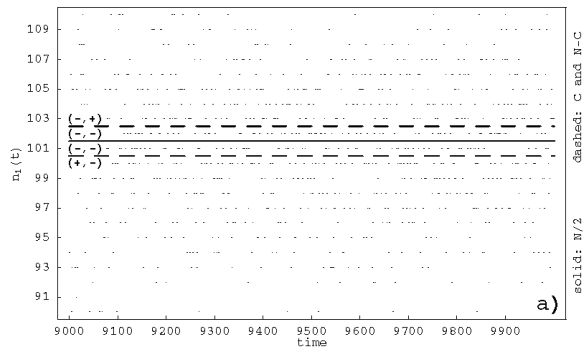

**Figure 2**

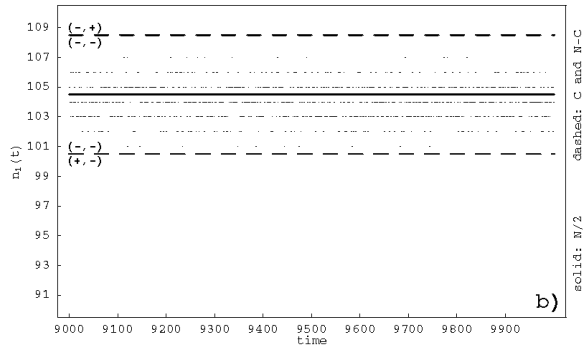

**Figure 3**

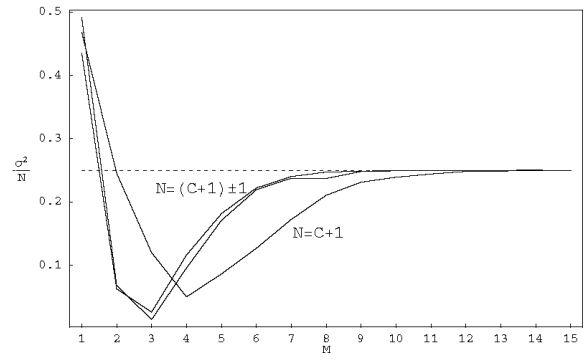

**Figure 4**

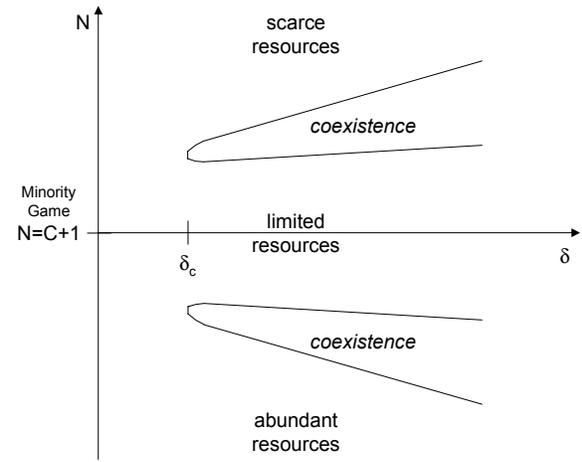

**Figure 5**